\documentclass{nature}
\usepackage{enumitem}
\usepackage{graphics}

\title{The Expanding Fireball of Nova Delphini 2013}

\author{G.~H.~Schaefer$^{1}$,
T.~ten Brummelaar$^{1}$,
D.~R. Gies$^{2}$, 
C.~D.~Farrington$^{1}$,
B.~Kloppenborg$^{2}$,  \\ 
O.~Chesneau$^{3}$, 
J.~D.~Monnier$^{4}$, 
S.~T.~Ridgway$^{5}$, 
N.~Scott$^{1}$, 
I.~Tallon-Bosc$^{6}$,
H.~A.~McAlister$^{2}$,   \\
T.~Boyajian$^{7}$, 
V.~Maestro$^{8}$,
D.~Mourard$^{3}$,
A.~Meilland$^{3}$,
N.~Nardetto$^{3}$, 
P.~Stee$^{3}$,
J.~Sturmann$^{1}$,  \\
N.~Vargas$^{1}$,
F.~Baron$^{2}$,  
M.~Ireland$^{9}$, 
E.~K.~Baines$^{10}$,
X. Che$^{4}$,
J.~Jones$^{2}$, 
N.~D.~Richardson$^{11}$, \\
R.~M.~Roettenbacher$^{4}$,
L.~Sturmann$^{1}$,  
N.~H.~Turner$^{1}$, 
P.~Tuthill$^{8}$, 
G.~van Belle$^{12}$, 
K.~von Braun$^{13}$, \\
R.~T.~Zavala$^{14}$,
D.~P.~K.~Banerjee$^{15}$,
N.~M.~Ashok$^{15}$,
V.~Joshi$^{15}$,
J.~Becker$^{4,16}$, and
P.~S.~Muirhead$^{17}$
}

\begin{document}

\vspace*{-1.03in} \noindent
 Published in {\it Nature} (515, 234--236, 13 November 2014, DOI: 10.1038/nature13834)

\vspace*{0.02in}

{\let\newpage\relax\maketitle}
%\maketitle

\begin{affiliations}
\item {The CHARA Array of Georgia State University, Mount Wilson Observatory, Mount Wilson, CA 91023, USA}
\item {Center for High Angular Resolution Astronomy and Department of Physics and Astronomy, Georgia State University, P.O. Box 5060, Atlanta, GA 30302, USA}
\item {Laboratoire Lagrange, UMR 7293, UNS-CNRS-OCA, Boulevard de l'Observatoire, CS 34229 - F 06304 Nice Cedex 4, France}
\item {Department of Astronomy, University of Michigan, 941 Dennison Bldg., Ann Arbor, MI 48109, USA}
\item {National Optical Astronomy Observatory, 950 North Cherry Ave., Tucson, AZ 85719, USA}
\item {Universite de Lyon; Universite Lyon 1, Observatoire de Lyon, 9 avenue Charles Andre, 69230 Saint Genis Laval; CNRS UMR 5574, Centre de Recherche Astrophysique de Lyon; Ecole Normale Superieure, France}
\item {Department of Astronomy, Yale University, New Haven, CT 06511, USA}
\item {Sydney Institute for Astronomy, School of Physics, University of Sydney, NSW 2006, Sydney, Australia}
\item {Research School of Astronomy \& Astrophysics, Australian National University, Canberra ACT 2611, Australia}
\item {Remote Sensing Division, Naval Research Laboratory, 4555 Overlook Avenue SW, Washington, DC 20375, USA}
\item {D\'{e}partement de Physique and Centre de Recherche en Astrophysique du Qu\'{e}bec (CRAQ), Universit\'{e} de Montr\'{e}al, C.P. 6128, Succ. Centre-Ville, Montr\'{e}al, QC H3C 3J7, Canada}
\item {Lowell Observatory, 1400 W. Mars Hill Rd., Flagstaff, AZ, 86001, USA}
\item {Max-Planck-Institute for Astronomy (MPIA), Konigstuhl 17, 69117 Heidelberg, Germany}
\item {United States Naval Observatory, Flagstaff Station, 10391 W.\ Naval Obs. Rd., Flagstaff, AZ 86001, USA}
\item {Astronomy and Astrophysics Division, Physical Research Laboratory, Navrangpura, Ahmedabad, Gujarat, 380009, India}
\item {Cahill Center for Astronomy and Astrophysics, California Institute of Technology, Pasadena, CA 91106, USA}
\item {Department of Astronomy, Boston University, Boston, MA 02215, USA}
\end{affiliations}

\setcounter{footnote}{34}

% Letters - no more than 4 pages.  Uninterupted page of text contains about 1300 words.  A typical Letter to Nature contains about 1500 words of text (excluding the first paragraph of Letters, figure legends,reference list, and the methods section if applicable) and four small display items (figures and/or tables) with brief legends.  A composite figure (with several panels) usually needs to take about half a page, equivalent to about 600 words, in order for the elements to be visible.  Maximum 30 references for Letters.

%%%%%%%%%%%%%%%%%%%%%%%%%%%%%%%%%%%%%%%%%%%%%%%%%%%%%%%%%%%%%%%%%%%%%%%
% pdftk nova_del_2013_nature.pdf figures_nature.pdf methods_nova.pdf cat output nova_del_2013_nature_2014jul11.pdf
%%%%%%%%%%%%%%%%%%%%%%%%%%%%%%%%%%%%%%%%%%%%%%%%%%%%%%%%%%%%%%%%%%%%%%%

%%%%%%%%%%%%%%%%%%%%%%%%%%%%%%%%%%%%%%%%%%%%%%%%%%%%%%%%%%%%%%%%%%%%%%%
% MAKE SURE TO ADD ARAS WEBSITE TO SHORE REFERENCE
% \url{http://www.astrosurf.com/aras/novae/Nova2013Del.html}
%%%%%%%%%%%%%%%%%%%%%%%%%%%%%%%%%%%%%%%%%%%%%%%%%%%%%%%%%%%%%%%%%%%%%%%

\begin{abstract}
A classical nova occurs when material accreting onto the surface of a white dwarf in a close binary system ignites in a thermonuclear runaway\cite{bode08,gehrz88}.  Complex structures observed in the ejecta at late stages\cite{slavin95,woudt09,chesneau12b} could result from interactions with the companion during the common envelope phase\cite{livio90,lloyd97}.  Alternatively, the explosion could be intrinsically bipolar, resulting from a localized ignition on the surface of the white dwarf \cite{orio93} or as a consequence of rotational distortion\cite{porter98,scott00}.  Studying the structure of novae during the earliest phases is challenging because of the high spatial resolution needed to measure their small sizes\cite{chesneau12a}.  Here we report near-infrared interferometric measurements of the angular size of Nova Delphini 2013, starting from one day after the explosion and continuing with extensive time coverage during the first 43 days.  Changes in the apparent expansion rate can be explained by an explosion model consisting of an optically thick core surrounded by a diffuse envelope.  The optical depth of the ejected material changes as it expands.  We detect an ellipticity in the light distribution, suggesting a prolate or bipolar structure that develops as early as the second day.  Combining the angular expansion rate with radial velocity measurements, we derive a geometric distance to the nova of 4.54 $\pm$ 0.59 kpc from the Sun.  

\end{abstract}

Nova Delphini 2013 (also known as V339 Del) was discovered\cite{nakano13} by Koichi Itagaki on 2013 Aug 14 Universal Time (UT) at 14:01.  We began an intensive observing campaign to measure the size of the nova with the CHARA Array\cite{tenbrummelaar05}, an optical/infrared interferometer located on Mount Wilson, California.  Our observations began within 15 hours of the discovery and within 24 hours of the detonation itself.  We measured the size of the expanding ejecta as the nova rose to peak brightness and continued monitoring it for a total of 27 nights between UT 2013 Aug 15 and Sep 26.

We measured the angular diameter of Nova Del 2013 by fitting a uniformly bright circle to the visibility amplitudes of the interference fringes recorded during each night (Extended Data Table 1).  We plot the expansion curve in Figure~1.  The dotted line shows a linear fit to the angular diameters during the first 27 days after the explosion.  The inset panel shows an apparent deceleration during the first four nights compared with lines of constant velocity.  During the last week, the measurements show a large jump in the effective size of the nova compared with an extrapolation of the linear fit.  For comparison, we show the visible and infrared light curves in Figure~2.  

The apparent changes in slope of the expansion curve can be explained by a two-component model consisting of an optically thick pseudo-photosphere surrounded by an optically thin halo.  We approximated the intensity distribution projected on the sky using a uniformly bright, circular core surrounded by a circular ring.   We fit the two-component model to the interferometric visibilities and minimized the total $\chi^2$ simultaneously across all nights.  We fixed the time of detonation at modified Julian day $t_0$ = MJD 56518.277, computed by extrapolating the first two pre-discovery photometric measurements\cite{wren13} back to the quiescent flux\cite{munari13_phot,deacon14}.  From our interferometric data, we measured an expansion rate of 0.156 $\pm$ 0.003 milli-arcseconds (mas) per day for the core diameter and a size ratio of 1.73 $\pm$ 0.02 between the outer ring and core.  We allowed the flux ratio between the ring and core to vary on a nightly basis.   

Figure~3 shows how the percentage of light from the ring changes over time.  During the first two nights, the core and ring have similar surface brightnesses.  Therefore, at the earliest times, the nova can be approximated by a single uniform disk component where the optically thick pseudo-photosphere extends to the outermost, fastest-moving layers of the ejecta.  After the peak in the visible light curve, the amount of flux in the ring drops, which indicates that the outer layers became optically thin and the pseudo-photosphere moved toward the inner, slower-moving, and denser layers.  Therefore, the apparent deceleration in the expansion curve during the first few nights is likely caused by a diminishing contribution from the outer layers.  Over the next few weeks, the absolute flux of both components decreased, however, the flux from the core dropped at a steeper rate compared with the ring.  Therefore, the percentage of light in the ring increased relative to the core.  These changes are consistent with the spectral evolution\cite{shore13b} of Nova Del 2013.  Optical spectra initially showed P Cygni absorption features during the optically thick fireball stage which disappeared after about five days.  Afterwards, strong emission lines began to dominate the spectrum, which are thought to form in the outer, optically thin layers\cite{gehrz88}.  The rising strength of the emission lines relative to the continuum was also seen in spectra at infrared wavelengths (Extended Data Fig.~1).  In the two-component model, we assumed a constant ratio between the size of the core and the halo.  In reality, changes in the optical depth of the ejecta are likely to be more complex and could be investigated further by fitting a physical model\cite{gill99,shore13tpyx} simultaneously to interferometric and spectroscopic data.  Understanding how the optical depth changes within the expanding ejecta impacts the physical interpretation of novae light curves\cite{shore13asym}.

During the last five nights, the two-component model shows an enhancement in the flux of the outer layers, with the ring contributing $\sim$ 68\% of the total light.  This coincides with an increase in the infrared flux from the nova\cite{taranova14} (see Fig.~2).  This suggests the formation of dust in the outer layers, perhaps in denser clumps within the ejecta\cite{gehrz88}, that contributes thermal emission.  Alternatively, as the ejecta expands, a recombination front might propagate through the material\cite{shore11}.  Near-infrared free-free emission from ionized gas the inner regions would decline and shift the effective size boundary to the outer layers.  This interpretation is consistent with the larger size of the effective photosphere at later dates, however, it does not explain the brightening in the near-infrared light curve.  The effective size of our two-component model is shown as the solid line in Figure~1; the model reproduces the changes in slope of the expansion curve.  

We reconstructed model-independent images of Nova Del 2013 using the MArkov Chain IMager\cite{ireland06} (MACIM) for data obtained on three nights during the first week with sufficient baseline coverage.  As shown in Figure~4, the images show a striking similarity to the two-component model, with an optically thick core surrounded by a halo of fainter emission.

Combining the angular expansion rate with the radial velocity of the ejected material provides a way to measure the distance\cite{gehrz88} to Nova Del 2013.  The velocities reported in the literature\cite{shore13b} range from $-600$ to $-2500$ km~s$^{-1}$, consistent with the expectation that lines form at different layers in the expanding atmosphere with a radial outflow velocity proportional to the distance from the white dwarf explosion site.  We selected the Si II $\lambda\lambda$ 6347, 6371 \AA ~absorption lines observed during the first week to represent the deeper layers, because they likely form in the dense, cooler gas immediately above the continuum forming layer.  Based on an analysis of spectra downloaded from the archive of the Astronomical Ring for Access to Spectroscopy\cite{shore13b}, we estimated the outflow speed near the continuum forming layer of $V_{\rm ejection} = 613 \pm 79$ km~s$^{-1}$.  Combining this velocity with the angular expansion rate, we derived a distance of $4.54 \pm 0.59$ kpc to Nova Del 2013.  Empirical evidence suggests that novae have prolate spheroid\cite{ford88} or bipolar\cite{shore13tpyx} shapes, therefore the true distance will depend on its geometry and inclination in the plane of the sky\cite{wade00}.  Assuming a probable inclination of 45$^\circ$ for Nova Del 2013 (S. Shore, personal communication, 16 April 2014), we estimate that these effects contribute to a systematic difference in the distance of less than 5\%.  Using the light curve decay times and reddening reported\cite{munari13_phot} for Nova Del 2013 (assuming an $R_V = 3.1$ extinction law), the maximum magnitude - rate of decline (MMRD) relation\cite{downes00} gives a distance ranging from 3.3 to 4.1 kpc, depending on the adoption of the linear or non-linear MMRD formulations.   The expansion parallax also confirms the distance of 4.2 kpc based on a spectral comparison\cite{shore13d} to nova OS Andromedae 1986.  

To investigate the development of asymmetric features in Nova Del 2013, we fit a uniformly bright ellipse to the visibilities on nights with coverage over a range of position angles on the sky (from two to nine days after the outburst; Extended Data Table~2).  For each night, the ellipse produces a significantly lower $\chi^2$ compared with the circular disk ($P < 0.01$ for most nights; see Methods).  The nova appears to be $\sim$ 13\% larger along the major axis compared with the minor axis, suggesting that the ejecta could be bipolar.  For the remaining nights with limited sky coverage, we fixed the axis ratio and position angle to their average values ($\theta_{\rm major}/\theta_{\rm minor}$ = 1.13 $\pm$ 0.07, PA = 128$^\circ$ $\pm$ 30$^\circ$) and solved for the size of the ellipse.  On average, we find that the circular disk diameters underpredict the mean angular diameter of the ellipse by $\sim$ 2\%.  Therefore, the assumption of circular symmetry in the two-component model should not strongly bias our results.

In addition to the visibility amplitudes that provide information on the size and shape of the source, interferometric closure phases obtained when combining the light from three or more telescopes indicate deviations from point symmetry (whether the brightness is the same when reflected through a point at the center of the source distribution).  The closure phases from the first two sets of data obtained with all six telescopes ($t$ = 3.0 and 5.0 days) are consistent with zero within the uncertainties, suggesting that the source was point-symmetric within the resolution limits.  By the third set of observations with all six telescopes ($t$ = 6.9 days; see Fig.~4), the closure phases rise steadily to $\sim 60^\circ$ (Extended Data Fig.~2), suggesting the detection of a point asymmetry in the brightness distribution.  If the outflow is bipolar, as suggested by our elliptical fits, then this could indicate a difference in the brightness between the two lobes (perhaps due to viewing angle) or, alternatively, the development of clumpy structures within the expanding material.

The ellipticity and closure phase asymmetries indicate that non-spherical structures developed as early as a few days after the outburst.  Along with results from previous interferometric studies of novae\cite{lane07a,chesneau07,chesneau11}, this suggests that novae explosions might be inherently bipolar or that the elliptical shape develops early during the common envelope phase.

%%%%%%%%%%%%%%%%%%%%%%%%%%%%%%%%%%%%%%%%%%%%%%%%%%%%%%%%%%%%%%%%%%%%%%%%
% Methods References - to get numbering correct

%tenbrummelaar05\cite{tenbrummelaar05}, tenbrummelaar13\cite{tenbrummelaar13}, monnier06\cite{monnier06}, bonneau06\cite{bonneau06}, boyajian13\cite{boyajian13}, crepp12\cite{crepp12}, skrutskie06\cite{skrutskie06}, munari13phot\cite{munari13_phot}

%tuthill06\cite{tuthill06}, markwardt09\cite{markwardt09}, press92\cite{press92}, wilson04\cite{wilson04}, herter08\cite{herter08}, cushing04\cite{cushing04}, muirhead11\cite{muirhead11}, vacca03\cite{vacca03}, das08\cite{das08} hjellming79\cite{hjellming79}

%%%%%%%%%%%%%%%%%%%%%%%%%%%%%%%%%%%%%%%%%%%%%%%%%%%%%%%%%%%%%%%%%%%%%%%%

%%%%%%%%%%%%%%%%%%%%%%%%%%%%%%%%%%%%%%%%%%%%%%%%%%%%%%%%%%%%%%%%%%%%%%%%%

\vspace{0.5cm}
\spacing{1.0}

\bibliography{nova_del_2013_nature}

%Bode, M. F., & Evans, A. 2008, Classical Novae (2nd ed.; Cambridge: Cambridge Univ. Press)

\spacing{1.5}

\begin{addendum}
%\item[Supplementary Information]is linked to the online version of the paper at www.nature.com/nature.
\item[Full Methods]and any associated references are available in the online version of the paper at \\
www.nature.com/nature.
 \item[Acknowledgements] We acknowledge the variable star observations from the AAVSO International Database contributed by observers worldwide and used in this research.  We thank O.~Garde and other members of the Astronomical Ring for Access to Spectroscopy for use of their archive of Nova Del spectra. We thank G.~J.~Schwarz, S.~N.~Shore, and F.~M.~Walter for discussions that helped us interpret the nova observations.  This material is based upon work supported by the National Science Foundation under Grant No.~AST-1009080.  The CHARA Array is funded by the National Science Foundation through NSF grants AST 0908253 and AST 1211129, and by Georgia State University through the College of Arts and Sciences.  This publication made use of data products from the Two Micron All Sky Survey, which is a joint project of the University of Massachusetts and the Infrared Processing and Analysis Center/California Institute of Technology, funded by the National Aeronautics and Space Administration and the National Science Foundation.
\item[Author Contributions] Observations with the CHARA Array were originally proposed by BK and DRG.  Modeling and interpreting the angular expansion curve and asymmetries were done by GHS, DRG, BK, TtB, OC, IT-B, and STR.  The CHARA data were reduced by TtB, JDM, OC, IT-B, DM, VM, CDF, NS, and GHS. The observations were planned and conducted by CDF, NS, NV, BK, DRG, TB, GHS, DM, AM, NN, PS, MI, VM, PT, JJ, NDR, RMR, GvB, KvB, and RTZ.  Observational setup and technical support were provided by JS, LS, NHT, and XC.  Administrative oversight and access to CHARA were provided by HAM and TtB.  Reconstructing and interpreting the nova images were done by GHS, FB, JDM, and BK.  Infrared magnitudes derived from CHARA data were computed by NS and GHS.  Infrared spectra were taken and reduced by DPKB, NMA, VJ, PSM, JB, and analyzed by DRG.  All authors discussed the results and commented on the manuscript.
\item[Author Information] Reprints and permissions information is available at
www.nature.com/reprints.  The authors declare no competing financial interests.
Correspondence and requests for materials should be addressed to G.H.S.~(schaefer@chara-array.org).
\end{addendum}

%%%%%%%%%%%%%%%%%%%%%%%%%%%%%%%%%%%%%%%%%%%%%%%%%%%%%%%%%%%%%%%%%%%%%%%%%
% Figure Captions
%%%%%%%%%%%%%%%%%%%%%%%%%%%%%%%%%%%%%%%%%%%%%%%%%%%%%%%%%%%%%%%%%%%%%%%%%

\clearpage

\noindent
%\begin{figure}
  \begin{center}
         \scalebox{1.0}{\includegraphics{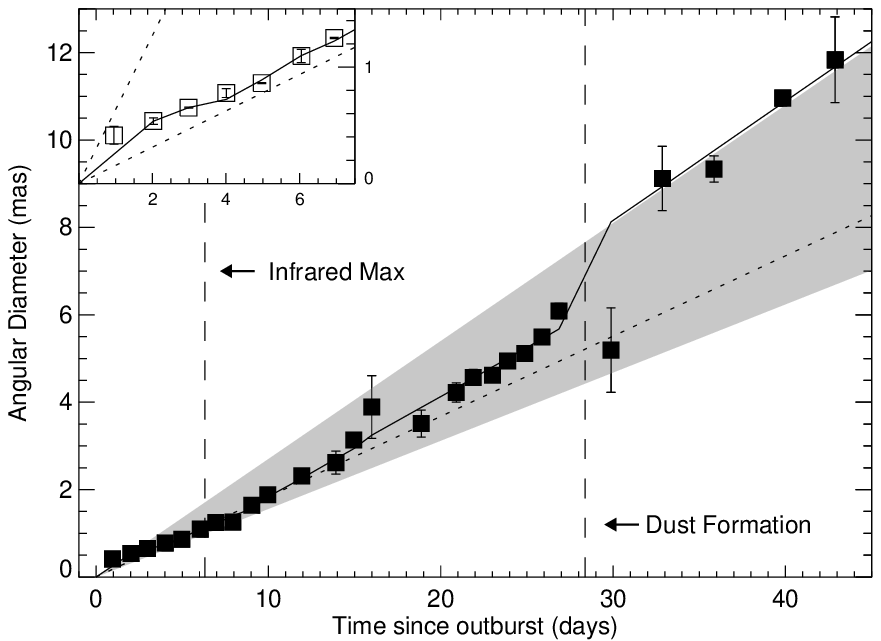}}
%        \scalebox{1.0}{\includegraphics{nova_expansion_hybrid_fring_all_rate0.156_scale1.73_inset_boot_udring.eps}}
 \end{center}
%       \caption{
{\bf Figure 1} -- Expansion curve of Nova Del 2013.  The angular diameters were measured by fitting a circular disk to the interferometric data.  The dotted line shows a linear fit for days 0--27.  The inset panel zooms in during the first week and shows dotted lines with velocities\cite{shore13b} of $613$ and $2500$ km~s$^{-1}$ at a distance of 4.54 kpc.  The apparent deceleration during the first week, along with the jump in size during the last week, can be explained by a two-component model consisting of a circular core surrounded by a ring where the flux ratio changes over time.  The grey region shows the expansion rate of the core (lower edge) and the ring (upper edge).  The solid line shows the effective size obtained by fitting the two-component model visibilities as a single circular disk. Error bars represent 1\,$\sigma$ uncertainties computed from a boostrap analysis.
%}
%\label{fig.nova}
%\end{figure}

\clearpage

\noindent
%\begin{figure}
  \begin{center}
         \scalebox{1.0}{\includegraphics{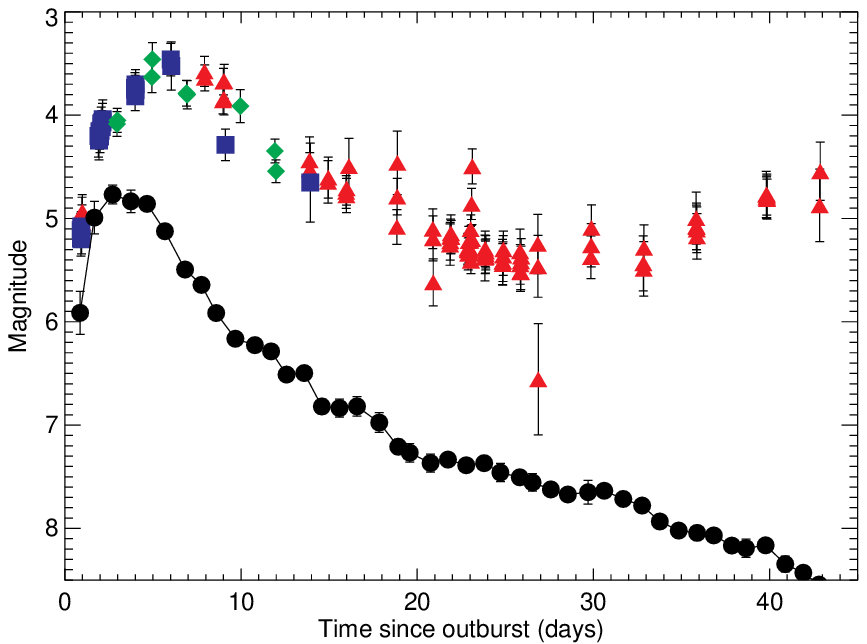}}
%        \scalebox{1.0}{\includegraphics{phot_cal_climb_classic_mirc_aavso_2014feb05_color.eps}}
  \end{center}
%       \caption{
{\bf Figure 2} -- Infrared light curve of Nova Del 2013.  The magnitudes and error bars were computed from the mean and standard deviation of counts recorded on the detector during each interferometric observation collected at the CHARA Array.  The data were obtained in the $H$-band using two different beam combiners (blue squares and green diamonds) and in the $K'$-band (red triangles). For comparison, we plot daily averages of photometric measurements in the $V$-band downloaded from the American Association of Variable Star Observers (AAVSO; black circles). The rise to peak brightness was slower in the infrared compared to the visible.
%}
%\label{fig.phot}
%\end{figure}

\clearpage

\noindent
%\begin{figure}
  \begin{center}
         \scalebox{1.0}{\includegraphics{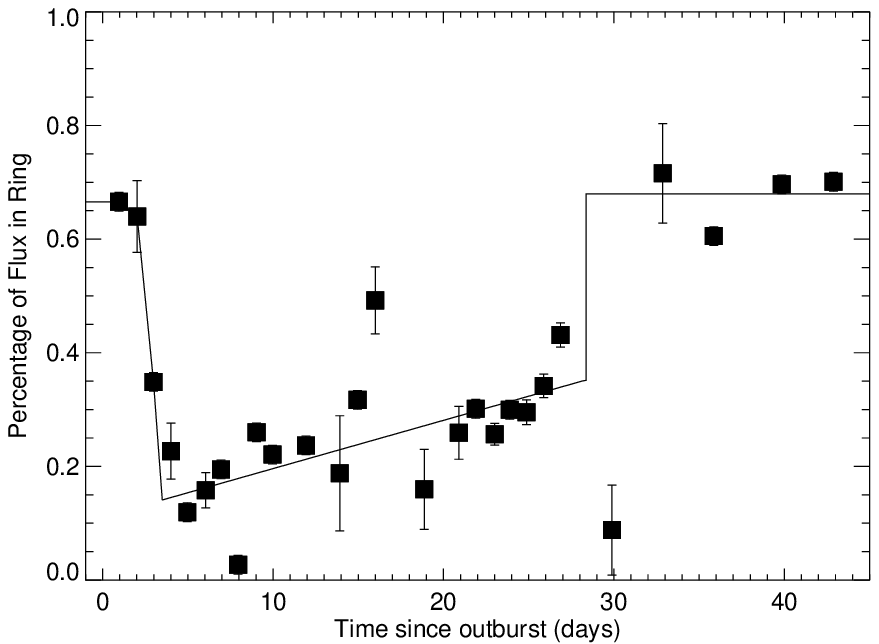}}
%    \scalebox{1.0}{\includegraphics{hybrid_fring_all_rate0.156_scale1.73_n3_20.eps}}
  \end{center}
%       \caption{
{\bf Figure 3} -- Changes in the flux ratio of the two-component model.  The solid line shows a fit where the ratio changes smoothly over time.  For the first three nights, the ring represents the outer boundary of the optically thick pseudo-photosphere. After the peak in the visible light curve, the flux in the ring drops.  On nights 4--27, the percentage of light from the ring increases at an approximately linear rate.  On the last four nights, the ring contributes an average of 68\% of the total light. Error bars represent 1\,$\sigma$ uncertainties derived from a least squares fit to the visibility data.
%}
%\label{fig.flux}
%\end{figure}

\clearpage

\noindent
%\begin{figure}
  \begin{center}
         \scalebox{1.0}{\includegraphics{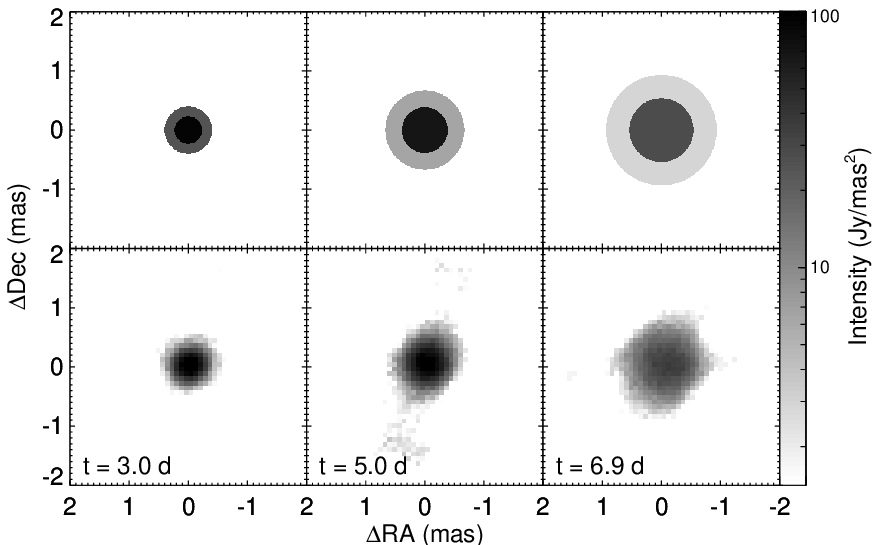}}
%          \scalebox{1.0}{\includegraphics{image_nova_cartoon_all_rate0.156_scale1.73_mirc_log_t_scale_label.eps}}
  \end{center}
%        \caption{
{\bf Figure 4} -- Model and reconstructed images of Nova Del 2013.  The two-component model consists of a circular core surrounded by a fainter ring (top row).  The images reconstructed using MACIM show a similar spatial structure (bottom row).  The data used in the reconstructions were obtained with the Michigan Infrared Combiner (MIRC) at the CHARA Array on UT 2013 Aug 17, 19, and 21 ($t$ = 3.0, 5.0, and 6.9 days after the outburst) and have good baseline coverage on the sky.  We scaled the flux by the infrared $H$-band magnitude measured on each night to show how the surface brightness changes.
%}
%\label{fig.images}
%\end{figure}
%%The resolution size of 0.5 mas corresponds to 2.27 AU at a distance of 4.54 kpc.  

%%%%%%%%%%%%%%%%%%%%%%%%%%%%%%%%%%%%%%%%%%%%%%%%%%%%%%%%%%%%%%%%%%%%%%%%%

\clearpage

\noindent
{\bf \large Methods}

\noindent
\begin{center}
{\bf Interferometric Observations with the CHARA Array} 
\end{center}

The CHARA Array\cite{tenbrummelaar05} is an optical/infrared interferometer located on Mount Wilson, California.  The array has six 1\,m telescopes arranged in a Y configuration with baselines ranging from 34 to 331\,m.  The longest baseline provides a spatial resolution of 0.5 mas at a wavelength of 1.6 $\mu$m.  We observed Nova Del 2013 in the infrared using the CLASSIC two-beam combiner$^{31}$, the CLIMB three-beam combiner$^{31}$, and the MIRC six-beam combiner$^{32}$.  The data were reduced using the standard reduction pipeline for each instrument.  Data products include the squared, normalized visibility amplitude of the interference fringes for the two-baseline combiners and the squared,normalized visibility amplitudes and closure phases for the instruments that combine the light from three or more telescopes (see Extended Data Figures 2--4).  Before and after each observation, we observed unresolved stars or stars with known angular diameters$^{33,34,35}$ to calibrate the interferometric measurements.  Extended Data Table 1 lists the UT date, modified Julian Day (MJD), time since the outburst ($t_0$ = MJD 56518.277), beam combiner, telescope configuration, filter, and calibrators used during each observation.  The calibrated data will be available though the Optical Interferometry Database developed by the Jean-Marie Mariotti Center (JMMC; http://www.jmmc.fr/oidb.htm).  On five nights we also collected data using two visible light beam combiners; these data will be discussed in a subsequent publication (O.C., D.M., P.T., V.M., I.T-B, D.P.K.B, E.Lagadec, M.I., N.N., A.M, P.S, R.T.Z., G.H.S, T.t.B., D.R.G., H.A.M., T.B., C.D.F., N.V., N.S., J.S., and L.S., manuscript in preparation).

In addition to the interferometric measurements, we also computed infrared magnitudes of Nova Del 2013 using the number of counts recorded on the detector.  We compared the counts from Nova Del 2013 with the calibrator stars observed immediately before and after.  We calibrated the photometry based on the 2MASS magnitudes$^{36}$ of the calibrators.  We present the photometry in Extended Data Table 3.  Figure 2 shows the infrared light curve of Nova Del 2013.  For comparison, we also plot daily averages of $V$-band photometric measurements downloaded from AAVSO$^{37}$.  The rise to peak brightness was slower in the infrared compared to the visible.  Based on a second-order polynomial fit to the $H$ and $K$-band measurements near the maximum ($<$ 4.3 mag), the near-infrared light curve reached a maximum of 3.61 mag at MJD 56524.6 ($t$ = 6.3 days).  The maximum brightness in the visible\cite{munari13_phot} occurred at an earlier time of MJD 56520.4 ($t$ = 2.1 days) at $V \sim$ 4.46 mag.

\noindent
\begin{center}
{\bf Correcting for the Effects of Line Emission} 
\end{center}

The emergence of strong emission lines in the infrared spectrum of Nova Del 2013 (such as Br$\gamma$ in the $K$-band) will broaden the central envelope of the interference fringe and create sidelobes farther out$^{38}$.  We investigated the effects of the emission lines on the calibration of the interferometric data obtained with the CLASSIC and CLIMB beam combiners.  The reduction code outputs the mean power spectrum for each target.  We fit a rectangular top-hat function to each power spectrum and calculated the mean center and width for the observations on each night.  By comparing the results from the nova with the calibrators, we found that the central wavelength of the filter remained constant over time.  We also found that the effective bandwidth of the filter remained the same during the first two weeks and then decreased by $\sim$ 14\% $\pm$ 3\% as the emission lines appeared in the nova spectra (see Extended Data Table~4).  Essentially, the emission lines put a larger fraction of the flux at the center of the wavelength region, thereby effectively decreasing the full-width at half-maximum (FWHM) of the filter.   The CLASSIC/CLIMB reduction software assumes a constant top-hat function for the width of the filter.  Therefore, when the effective bandwidth decreases, the visibility is overestimated and the angular size of the nova is underestimated.  

To correct for this effect, we multiplied the CLASSIC/CLIMB visibilities by the ratio of the mean effective width of the nova observations relative to the calibrators measured on each night.  We added the uncertainty in the bandwidth ratio in quadrature with the uncertainties in the visibilities reported by the reduction code.  We applied this correction starting on UT 2013 Aug 28, when the effects of the emission lines became measurable.  During the first two weeks, when the contribution from the emission lines had a negligible effect on the calibration, we assumed the default bandwidth used in the reduction code, but included the average uncertainty in the nova-to-calibrator bandwidth ratio of $\pm 0.164$ in the error budget so that all of the CLASSIC and CLIMB visibilities would have similar weights during the global two-component fit.  For the uniform disk fits to the data from each night, the bandwidth correction led to the measurement of the angular size of the nova being larger at the $\sim$ 5\% level after the emission lines developed.

\noindent
\begin{center}
{\bf Uniform Disk and Uniform Ellipse Fits} 
\end{center}

For each individual night of CHARA data, we measured the angular diameter of Nova Del 2013 by fitting a uniformly bright, circular disk to the visibilities and performing a Levenberg-Marquardt least-squares minimization using the IDL mpfit package$^{39}$.  The uniform disk diameters ($\theta_{\rm UD}$), reduced $\chi^2_\nu$ (where $\nu$ is the number of degrees of freedom), and number of visibility measurements $N(V^2)$ on each night are listed in the last three columns of Extended Data Table~1.  

For five nights with sufficient baseline coverage over a range of position angles, we fit a uniformly bright ellipse to the visibility data.  The fitted parameters include the size of the major and minor axes and position angle of the major axis ($\theta_{\rm major}$, $\theta_{\rm minor}$, PA); the values are given in Extended Data Table~2.  For four out of the five nights, performing an F-test indicates that the uniform ellipse produces a significant improvement in the $\chi^2$ compared with a uniform circle, at a significance level of 0.01.  On UT 2013 Aug 16, the improvement is only at the 0.10 significance level.

For both the uniform disk and uniform ellipse fits, we determined uncertainties in the parameters using a bootstrap approach$^{40}$.  During each iteration, we created a synthetic sample of visibility measurements from our original set of measurements by randomly selecting the same total number of visibilities, with replacement.  Therefore, in the synthetic sample, some measurements are repeated and others are left out.  We applied normally distributed uncertainties to the synthetic measurements (using the measured uncertainties from the original data).  We then fit a uniform circle or uniform ellipse to the synthetic data set to determine the best fit parameters for that iteration.  We performed 1000 iterations and computed uncertainties based on the standard deviation of the parameter distributions.  These uncertainties are reported in Extended Data Tables~1 and~2.

We expected the size of the nova to vary over the course of the night.  The maximum length of observing time was $\sim$ 6 hours on UT 2013 Aug 16 and Sep 6.  Given an expansion rate of 0.156 mas/day, we would expect the size of the nova to change by $\sim$ 0.04 mas over the course of the observations.  For the three nights with the smallest uncertainties on the angular diameter measurements ($\pm$ 0.005 mas on UT 2013 Aug 17, 19, and 21), the length of observing time ranged from 30 to 45 minutes, so the change in size is expected to be $<$ 0.005 mas.  For all other nights, the length of observing time varied from $\sim$ 30 minutes to 3 hours, corresponding to a change in size of 0.003 to 0.02 mas.  For these nights, the quoted uncertainties on the angular diameters in Extended Data Table 1 are larger than $\pm$ 0.02 mas, which should account for any changes in size over the course of the night.  For the two-component model described in the next section, we fit a dynamic model directly to the visibilities versus time, so it accounts for changes in size over the course of the night.

\noindent
\begin{center}
{\bf Two-Component Model} 
\end{center}

The two-component model consists of a uniformly bright, circular disk surrounded by a uniformly bright ring.  The angular diameter of the central uniform disk ($\theta_{\rm UD}$) and the outer diameter of the ring ($\theta_{\rm ring}$) are given by:
\begin{equation}
\theta_{\rm UD}(t) = \frac{d\theta}{dt} \, (t - t_0)
\end{equation}
\begin{equation}
\theta_{\rm ring}(t) = C_{\rm ring} \, \theta_{\rm UD}(t)
\end{equation}
where $\frac{d\theta}{dt}$ is the angular expansion rate of the central core and $C_{\rm ring}$ is the ratio of the outer ring diameter relative to the uniform disk diameter.

We determined the two global parameters ($\frac{d\theta}{dt}$ and $C_{\rm ring}$) by minimizing the cumulative $\chi^2$ between the measured and model visibilities across all nights.  To do this, we searched through a grid of values for the expansion rate and size ratio and calculated the size of each component during the times of observation.  Then, for each night of observation, we performed a Levenberg-Marquardt least-squares minimization using the IDL mpfit package$^{39}$ to solve for the best-fitting flux ratio between the ring and the core component.  

There are degeneracies between the overall size of the nova and the flux ratio of the two components; a smaller expansion rate can be compensated by placing more flux into the outer component.  Therefore, we followed a two-step procedure in fitting the expanding two-component model simultaneously to all nights.  Initially, we allowed the flux in the ring relative to the core to vary each night, but with a limit set so that the surface brightness of the ring would not exceed that of the core.  The initial restriction prevented a large increase in the fitted brightness of the ring at later nights.  Using the grid-search described above, we determined an expansion rate of 0.156 $\pm$ 0.003 mas per day for the core diameter and a size ratio of 1.73 $\pm$ 0.02 between the outer ring diameter and the core diameter.  We then fixed the expansion rate and size ratio and solved for the ring-to-core flux ratio for each night after removing the restriction on the surface brightness on all but the first three nights (we would not expect a clearing of the central region at such early times).  This allows for a moderate brightening of the ring at later times.  

In Figure~1,  the solid line shows the effective angular size of the two-component model.  This curve was generated by assuming that the percentage of light from the ring varies smoothly over time (solid line in Fig.~3) and computing the model visibilities for each night based on the expansion rate and size ratio.  We then fit a single uniform, circular disk to the model visibilities from each night to compute the effective size.  Extended Data Figure~5 shows how the size and flux ratios of the two-components change during the nights of observation.

\noindent
\begin{center}
{\bf Emission Line Flux from Infrared Spectroscopy}
\end{center}

To investigate how the emission line flux changed in the infrared, we used low dispersion spectra obtained with the TripleSPEC spectrograph$^{41,42}$ on the 200-inch Hale Telescope at Palomar Observatory on UT 2013 Aug 20 and 23.  We reduced the spectra using the Spextool reduction program modified for TripleSpec$^{43}$, which includes routines to correct for detector artifacts$^{44}$.  We calibrated and removed the telluric lines in the nova spectrum using the xtellcor program$^{45}$, which compares a high signal-to-noise spectrum of Vega to observations of a nearby A0 star, SAO 88500, taken immediately after each observation of the nova.  We also used infrared spectra$^{46}$ obtained at Mount Abu, India$^{47}$ on UT 2013 August 28 and 29, and September 8 and 20.

We collected these spectra into two regions covering the wavelength range of the CHARA CLASSIC/CLIMB $H$-band and $K'$-band filters and normalized the flux to unity at 1.65 and 2.2 $\mu$m, respectively.  The $H$-band is dominated by the upper level transitions of the hydrogen Brackett series plus a number of C I lines, and the $K'$-band is dominated by Br$\gamma$, He I $\lambda$ 2.0585 $\mu$m, and some C I lines$^{47}$.

To estimate the flux from the continuum, we made spline fits to the lower envelope of each spectrum.  We then summed both the continuum flux and the emission flux (total minus continuum) across the wavelength bands to form a ratio of the integrated emission line flux relative to the integrated continuum flux.  As shown in Extended Data Figure~1, the emission strength grew over the time-frame of the CHARA observations of Nova Del 2013 and began to decrease in the final spectrum.  These changes are similar to the changes we measured in the effective bandwidth of the interferometric observations.  

The strengthening in the emission lines helps explain why the expansion curve of Nova Del 2013 became steeper after the first week.  The emission lines form over a greater radial extent in the expanding envelope than does the continuum from the pseudo-photosphere.  Thus, if the outflow follows a Hubble-type relation with the velocity proportional to radial distance$^{48}$, then the emission lines sample parts of the envelope that were expanding faster than the photosphere.  We interpolated linearly between the spectroscopic measurements to obtain the ratio of the emission line to continuum flux at the times of the CHARA observations and used this as the initial estimate of the flux ratio in the halo ring relative to the uniform disk core in the two-component fit.

The drop in the emission to continuum flux ratio after the fourth week can be attributed to the onset of dust formation.  This can be seen in the $K$-band spectra, where the continuum slopes down in the first spectra and becomes flat in the last spectrum.  This is consistent with flux from a dusty component that contributes more to the continuum and increases with wavelength.  Our two-component model based on the interferometric observations suggests that if dust formation occurred, that it happened over a larger spatial scale than that of the expanding photosphere.  

We also made measurements of the flux weighted central wavelength of each infrared spectrum.  These were essentially constant to within a percent or so during these observations, consistent with our results from the analysis of the fringe power spectra.

\noindent
\begin{center}
{\bf Measuring the Radial Velocity of the Optically Thick Core}
\end{center}

In order to compare the angular expansion rate with the radial velocity of the nova outflow, we needed to decide which features in the optical spectra are most representative of the kinematics of the near-infrared continuum.   The velocities reported in the literature\cite{shore13b} range from $-600$ to $-2500$ km~s$^{-1}$, consistent with the expectation that lines form at different layers in the expanding atmosphere with a radial outflow velocity proportional to distance from the white dwarf explosion site.  The optical depth unity boundary for the near-infrared continuum must form at higher densities deep inside the outflow (compared to the P Cygni lines and emission lines that form further out in lower density gas).  The absorption lines observed during the first week are probably the most representative of the deeper layers, because they form in the dense, cooler gas immediately above the continuum forming layer.  We selected the Si~II $\lambda\lambda$ 6347, 6371 \AA ~absorption lines for measurement because they were observed throughout the first week and are relatively free of blending and interference from Earth's atmospheric lines and nova emission lines.  

We downloaded six spectra with a high resolving power ($R$ = 10,000) obtained over the first week by Olivier Garde from the archive of the Astronomical Ring for Access to Spectroscopy\cite{shore13b} and transformed them to a unit continuum on a heliocentric wavelength grid.  We measured radial velocities for the Si~II lines by cross-correlation with a model spectrum ($T_{\rm eff} = 8000$~K, $\log g = 2.0$, solar metallicity) based upon ATLAS atmospheres from Robert Kurucz.  The mean velocity and standard deviation from this sample (MJD 56519.8 -- 56524.8) are $-598 \pm 51$ km~s$^{-1}$.  The local standard of rest radial velocity at the probable distance of the nova is $+15$ km~s$^{-1}$, so the implied outflow velocity is $-613 \pm 51$ km~s$^{-1}$. However, we do not know the peculiar velocity of the nova relative to its local standard of rest, so we applied a representative velocity dispersion of 60 km~s$^{-1}$ for white dwarfs to revise the uncertainty in our final estimate of the outflow speed near the continuum forming layer, $V_{\rm ejection} = 613 \pm 79$ km~s$^{-1}$. Note that this is strictly an upper limit for the velocity of the continuum forming layers because the absorption lines form above the continuum in a somewhat faster outflow.  Additionally, the estimate reflects the geometry of the outflow which is dependent on the axial inclination for a bipolar ejection.  For example, if the absorption lines formed in a spherically expanding photosphere, then we might expect to observe line boundaries ranging from zero (flux from the limb) to $-V_{\rm outflow}$ (flux from the center), with a mean velocity between these limits.  However, the observed Si~II lines are rather narrow (FWHM $\sim$ 280 km~s$^{-1}$), contrary to this prediction.  On the other hand, for a bipolar outflow the absorption velocities may be much more restricted to the radial ejection speed along the line of sight.  Consequently, until future observations reveal details of the geometry of the nova remnant, we will simply assume that the measured radial velocity is the same as the radial outflow speed in our direction and that this is also the same as the transverse expansion speed as observed in the plane of the sky.

\noindent
{\bf \large Method References}

\noindent
31. ten Brummelaar, T. A., et al. The CLASSIC/CLIMB Beam Combiner at the CHARA Array. {\it Journal of Astronomical Instrumentation} {\bf 2}, 97, 1--20 (2013).

\noindent
32. Monnier, J. D., et al. Michigan Infrared Combiner (MIRC): commissioning results at the CHARA Array. {\it Proc. SPIE} {\bf 6268}, 62681P, 1--11 (2006).

\noindent
33. Bonneau, D., et al. SearchCal: a virtual observatory tool for searching calibrators in optical long baseline interferometry. I. The bright object case. {\it Astron. Astrophys.} {\bf 456}, 789--789 (2006).

\noindent
34. Boyajian, T.~S., et al. Stellar Diameters and Temperatures. III. Main-sequence A, F, G, and K Stars: Additional High-precision Measurements and Empirical Relations. {\it Astrophys. J.} {\bf 771}, 40, 1--31 (2013).

\noindent
35. Crepp, J.~R., et al. The Dynamical Mass and Three-dimensional Orbit of HR7672B: A Benchmark Brown Dwarf with High Eccentricity. {\it Astrophys. J.} {\bf 751}, 97, 1--14 (2012).

\noindent
36. Skrutskie, M. F., et al. The Two Micron All Sky Survey (2MASS). {\it Astron. J.} {\bf 131}, 1163--1183 (2006).

\noindent
37. Henden, A.A., The AAVSO International Database, http://www.aavso.org (2013).

\noindent
38. Tuthill, P., et~al. Double-Fourier spatio-spectral decoding. In {\it Advances in Stellar Interferometry}, {\it SPIE Conference Series}, {\bf 6268}, 62680X, 1--9 (2006).

\noindent
39. Markwardt, C.~B. Non-linear Least-squares Fitting in IDL with MPFIT.  In {\it Astronomical Data Analysis Software and Systems XVIII}, {\it Astronomical Society of the Pacific Conference Series}, {\bf 411}, 251--254 (2009).  http://purl.com/net/mpfit

\noindent
40. Press, W. H., Teukolsky, S. A., Vetterling, W. T., \& Flannery, B. P. {\it Numerical Recipes in C} (Cambridge Univ. Press, New York, NY 10011, USA, 1992)

\noindent
41. Wilson, J. C., et al. Mass producing an efficient NIR spectrograph. In {\it Ground-based Instrumentation for Astronomy}, {\it SPIE Conference Series}, {\bf 5492}, 1295--1305 (2004).

\noindent
42. Herter, T. L., et al. The performance of TripleSpec at Palomar. In {\it Ground-based and Airborne Instrumentation for Astronomy II}, {\it SPIE Conference Series} {\bf 7014}, 70140X, 1--8 (2008).

\noindent
43. Cushing, M. C., Vacca, W. D. \& Rayner, J. T. Spextool: A Spectral Extraction Package for SpeX, a 0.8-5.5 Micron Cross-Dispersed Spectrograph. {\it PASP} {\bf 116}, 362--376 (2004).

\noindent
44. Muirhead, P. S., et al.\ Precise Stellar Radial Velocities of an M Dwarf with a Michelson Interferometer and a Medium-Resolution Near-Infrared Spectrograph. {\it Publ. Astron. Soc. Pacif.} {\bf 123}, 709--724 (2011). 

\noindent
45. Vacca, W. D., Cushing, M. C. \& Rayner, J. T. A Method of Correcting Near-Infrared Spectra for Telluric Absorption. {\it Publ. Astron. Soc. Pacif.} {\bf 115}, 389--409 (2003).

\noindent
46. Banerjee, D. P. K., Ashok, N. M., Joshi, V. \&  Evans, A. Ongoing near-infrared observations of V339 Del (Nova Del 2013). {\it Astron. Telegr.} {\bf 5404}, 1 (2013).

\noindent
47. Das, R. K., Banerjee, D. P. K., Ashok, N. M. \& Chesneau, O. Near-infrared studies of V1280 Sco (Nova Scorpii 2007). {\it Mon. Not. R. Astron. Soc.} {\bf 391}, 1874--1886 (2008).

\noindent
48. Hjellming, R. M., Wade, C. M., Vandenberg, N. R. \& Newell, R. T. Radio emission from nova shells. {\it Astron. J.} {\bf 84}, 1619--1631 (1979).

\clearpage

\noindent
{\bf Extended Data Table 1} -- Journal of observations and angular diameters measurements of Nova Del 2013.  The angular diameters were computed by fitting a circular uniform disk to the visibilities from each night. \\
\vspace{-2.0cm}
\begin{center}
\includegraphics{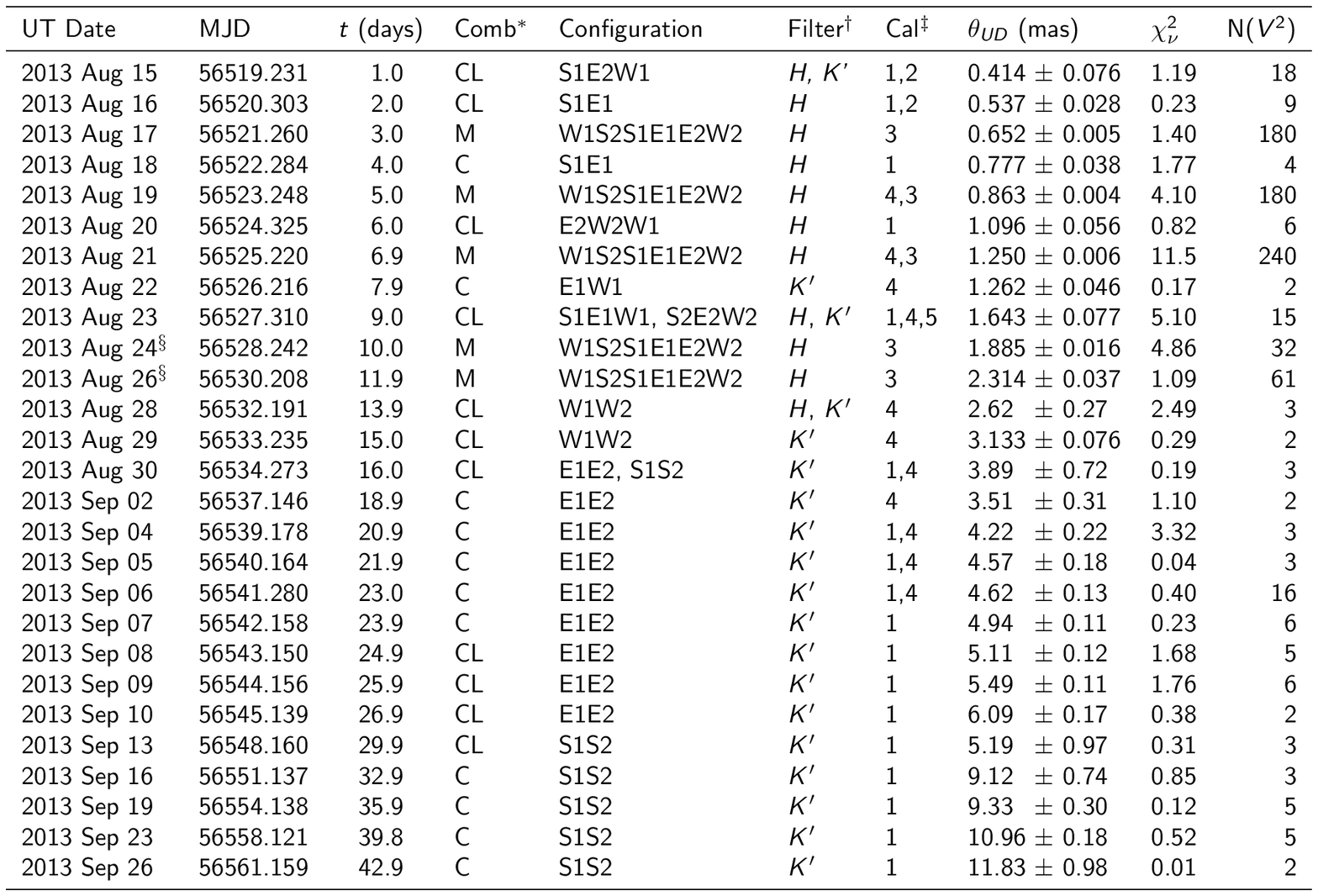}
\end{center} 
{\bf Footnotes to Extended Data Table 1} --
\vspace{-0.8cm}
\begin{description}[style=multiline,leftmargin=0.4cm,font=\normalfont]
\itemsep0em
\item[$^{*}$] Beam Combiner Codes: C = CLASSIC, CL = CLIMB, M = MIRC
\item[$^{\dag}$] CLASSIC/CLIMB $H$: $\lambda =$ 1.6731 $\mu$m, $\Delta\lambda = $ 0.2854 $\mu$m; CLASSIC/CLIMB $K'$: $\lambda =$ 2.1329 $\mu$m, $\Delta\lambda = $  0.3489 $\mu$m;  MIRC $H$: $\lambda$ = 1.5--1.7 $\mu$m with $\Delta \lambda \sim$ 0.034 $\mu$m for each of the 8 spectral channels.
\item[$^{\ddag}$] Calibrator Codes: 1 -- HD 190993 ($\theta = 0.161 \pm 0.011$ mas [33]), 2 -- HD 196740 ($\theta = 0.180 \pm 0.012$ mas [33]),  3 -- HD 206860 ($\theta = 0.530 \pm 0.015$ mas [34]), 4 -- HD 190406 ($\theta = 0.584 \pm 0.010$ mas [35]),  5 -- HD 195810 ($\theta = 0.301 \pm 0.021$ mas [33])
\item[$^{\S}$] Only E2W1W2 and S1S2 were recorded on UT 2013 Aug 24.  Only W1W2, E1E2, S1S2, and E2W2 were recorded on UT 2013 Aug 26.
\end{description}

%%\clearpage

\noindent
{\bf Extended Data Table 2} -- Uniform ellipse models of Nova Del 2013.
\begin{center}
\includegraphics{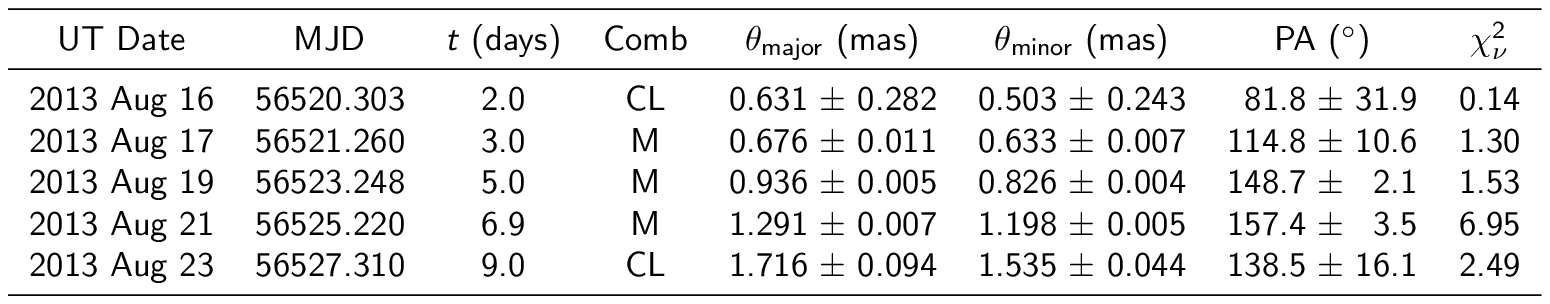}
\end{center} 

\clearpage

\noindent
{\bf Extended Data Table 3} -- Infrared Magnitudes of Nova Del 2013.
\begin{center}
\includegraphics{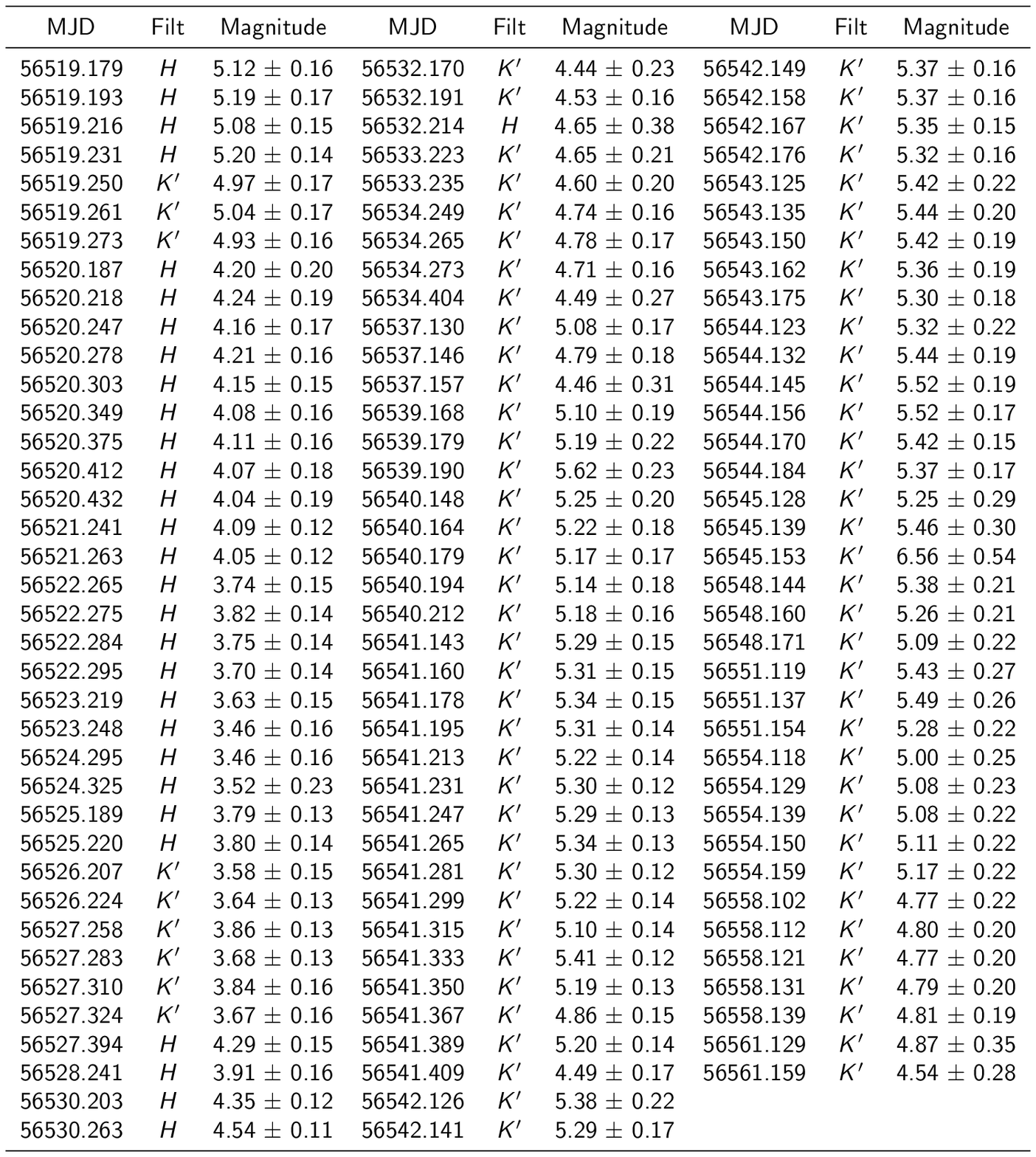}
\end{center} 

\clearpage

\noindent
{\bf Extended Data Table 4} -- Effective bandwidth of the CLASSIC/CLIMB observations.  We present the ratio of the effective bandwidth of the interferometric observations of Nova Del 2013 compared with the calibrator stars.
\begin{center}
\includegraphics{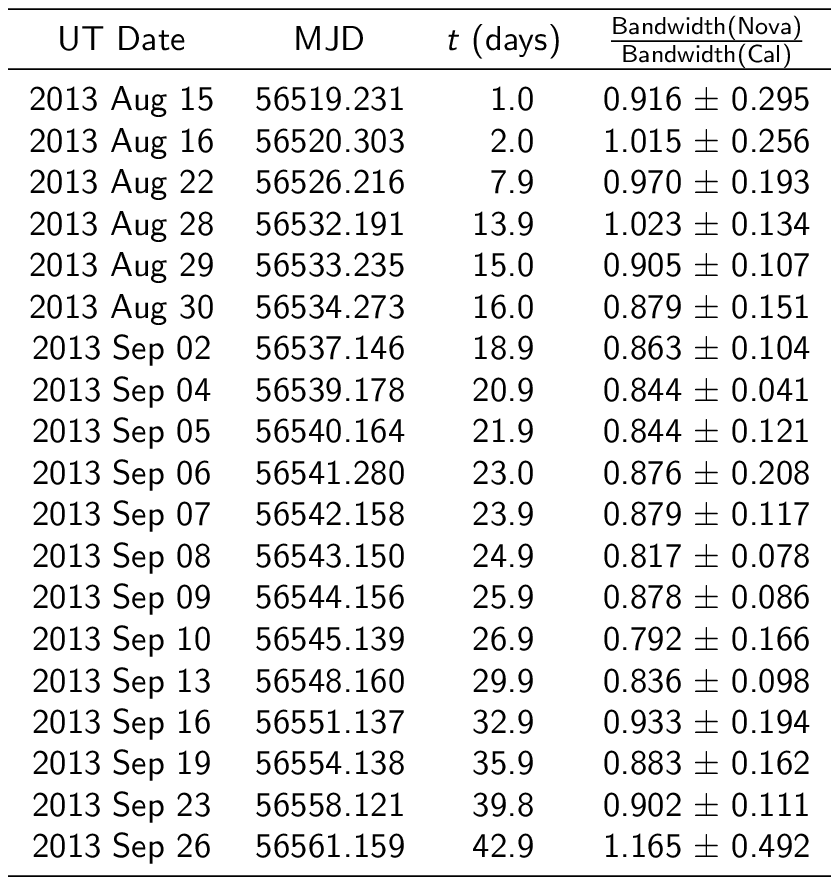}
\end{center} 

\clearpage

\noindent
\begin{center}
\includegraphics{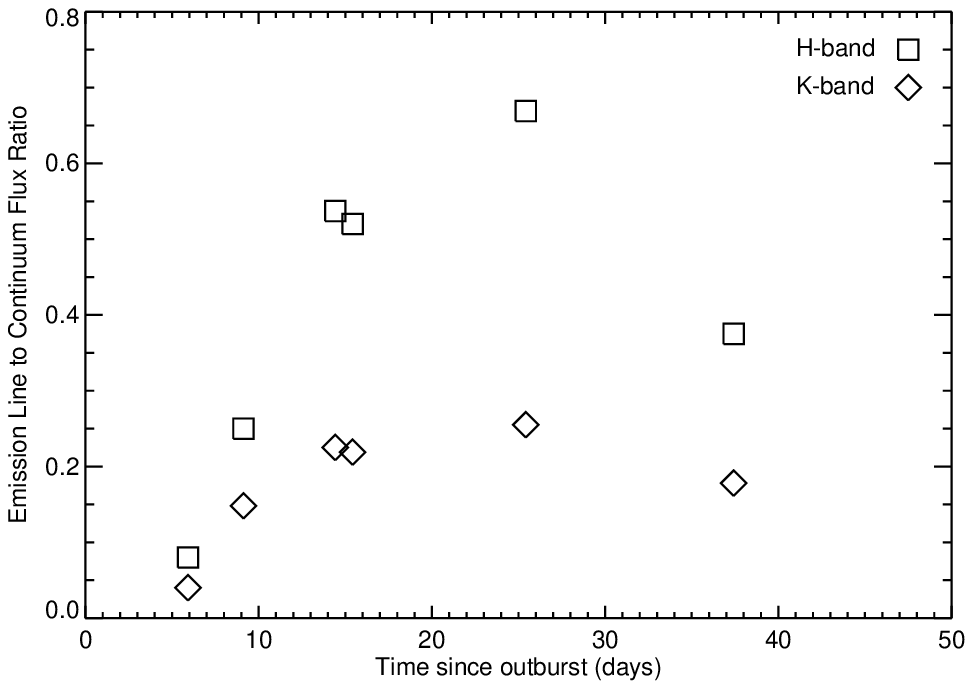}
\end{center} 
{\bf Extended Data Figure 1} -- Integrated emission line to continuum flux ratios measured from infrared spectroscopy.  The squares represent the $H$-band ratios while the diamonds represent the $K'$-band ratios.  The rise in the emission line flux is consistent with an increasing contribution from optically thin emission.  The down turn in the curve during the last measurement is likely caused by a rising contribution of the continuum from the formation of dust.

\clearpage

\noindent
\begin{center}
\includegraphics{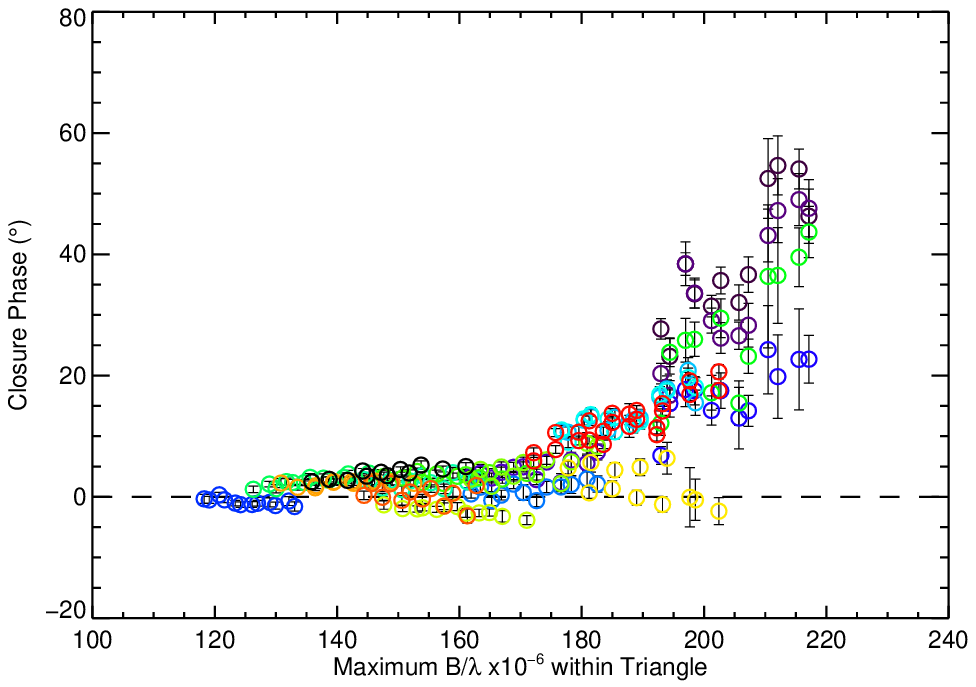}
\end{center} 
{\bf Extended Data Figure 2} -- Closure phases measured with MIRC on UT 2013 Aug 21. The measurements are plotted against the length of the maximum baseline ($B$) for the group of three telescopes that were used to form the closure phase.  The non-zero closure phases indicate an asymmetry in the light distribution that is not point symmetric.  The colors of the symbols are used to differentiate the measurements from each grouping of three telescopes in the array.  The closure phases are measured in eight wavelength channels.  Error bars represent 1\,$\sigma$ measurement uncertainties.

\clearpage

\noindent
\begin{center}
\includegraphics{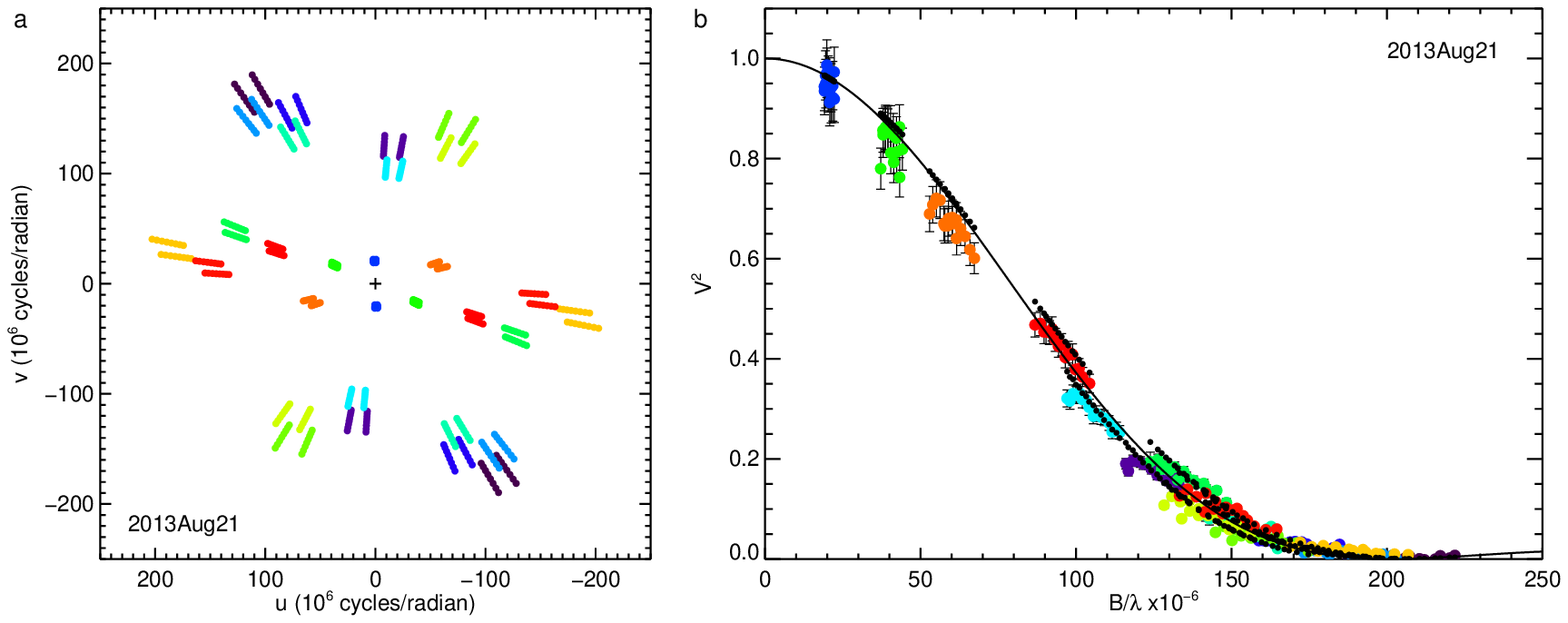}
\end{center} 
{\bf Extended Data Figure 3} -- Example of MIRC observations obtained on UT 2013 Aug 21.  The light is dispersed over 8 wavelength channels in the $H$-band.  {\bf a}, Coverage of interferometric baselines projected on the plane of the sky in right ascension (RA) and declination (Dec.) in units of spatial frequencies ($u = B_x/\lambda$, $v = B_y/\lambda$).  {\bf b}, Squared, normalized visibility amplitude measurements, color coded to match the baselines on the left. The solid line shows the best fit uniform disk model.  The small black dots show the best fit uniform ellipse model.  Error bars represent 1\,$\sigma$ measurement uncertainties.

\clearpage

\noindent
\begin{center}
\includegraphics{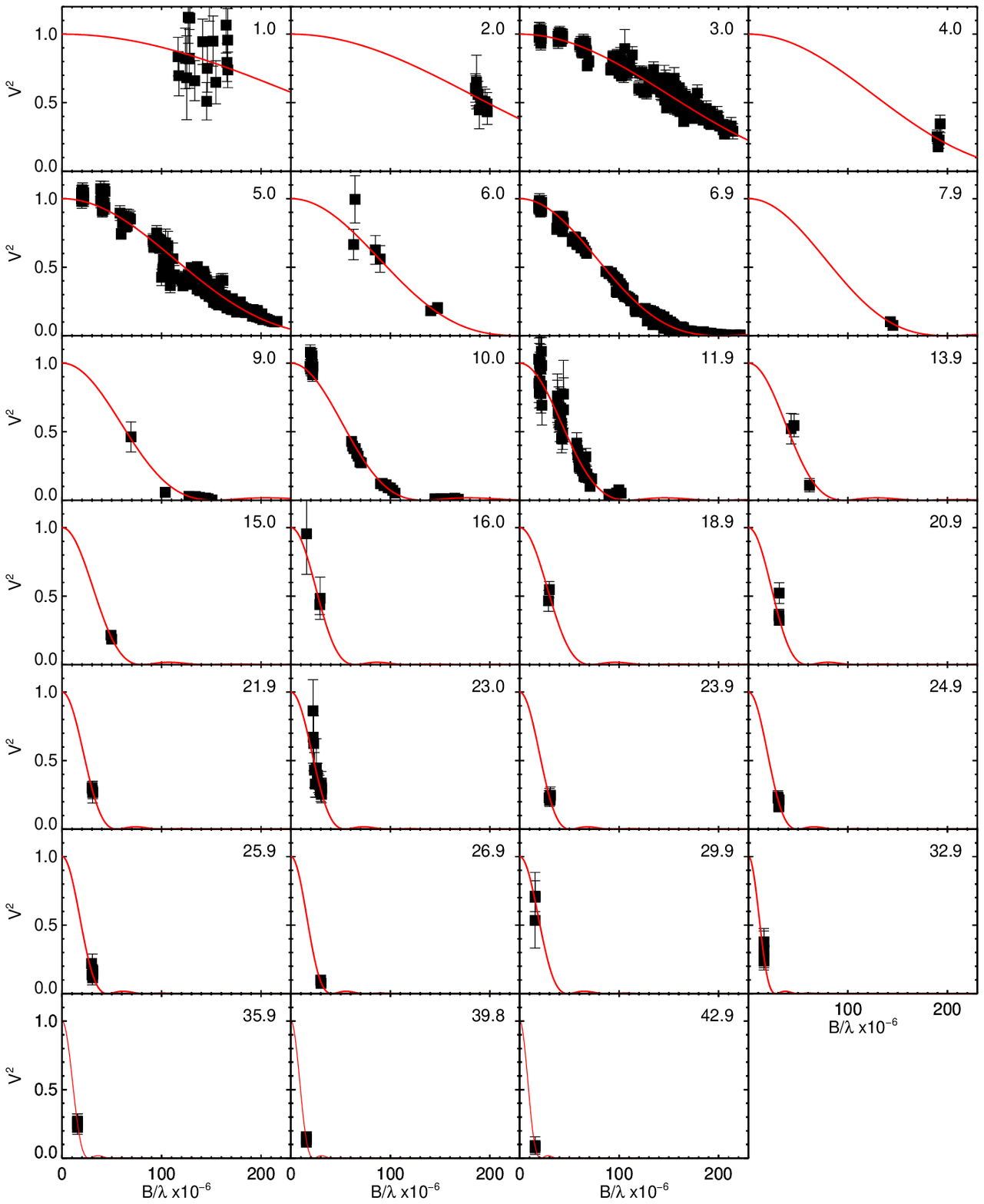}
\end{center} 
{\bf Extended Data Figure 4} -- Interferometric visibilities of Nova Del 2013 measured with the CHARA Array.  The red line shows the best-fitting model for a uniformly bright, circular disk.  The time since the explosion (in days) is indicated in the upper right corner of each panel.  The measurements were obtained with the CLASSIC, CLIMB, and MIRC beam combiners (see Extended Table 1).  Error bars represent 1\,$\sigma$ measurement uncertainties.

\clearpage

\noindent
\begin{center}
\scalebox{0.9}{\includegraphics{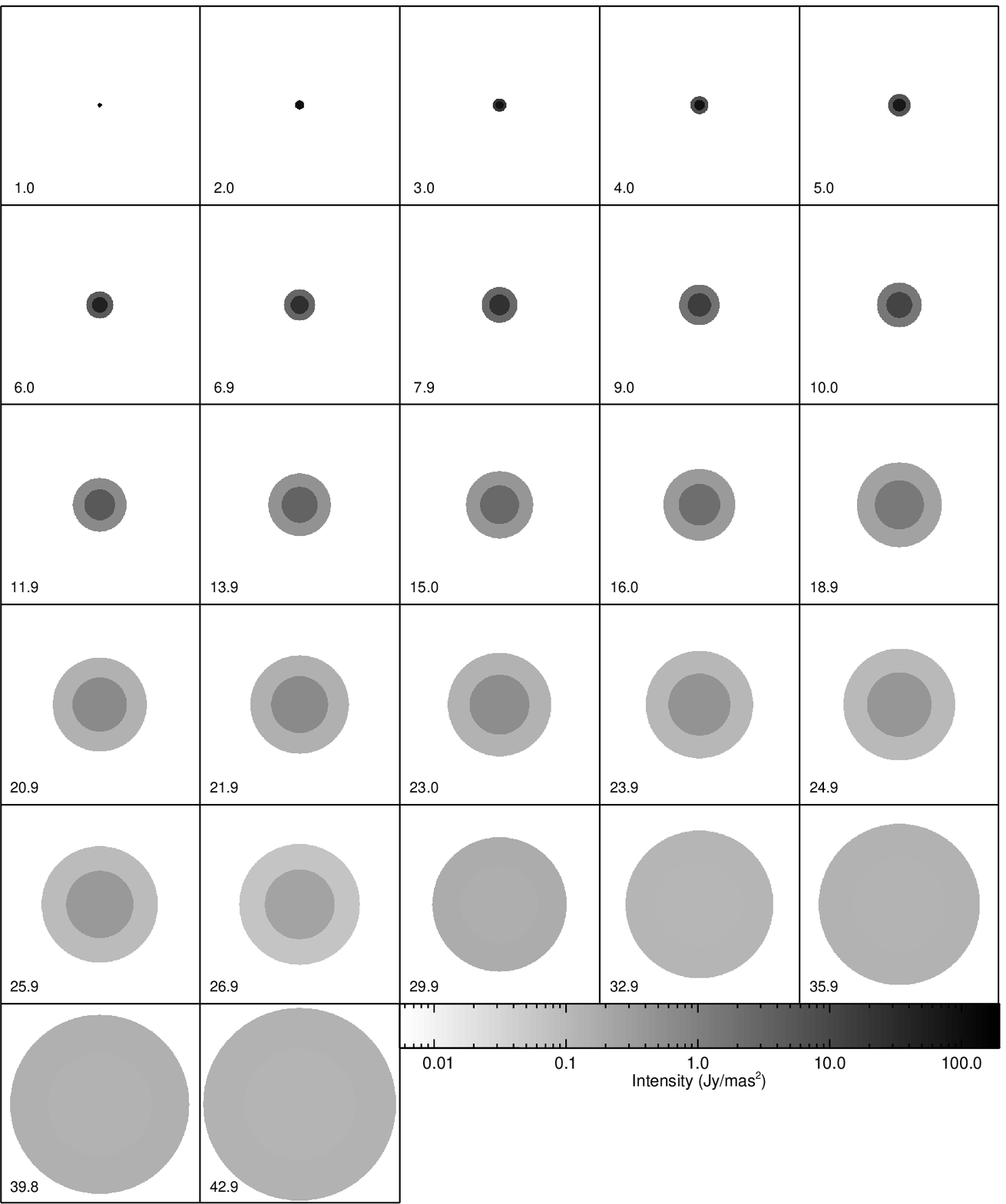}}
%\scalebox{0.9}{\includegraphics{ext_data_fig5.eps}}
%\scalebox{0.9}{\includegraphics{image_nova_cartoon_all_rate0.156_scale1.73_hybrid_log_t_scale.eps}}
\end{center} 
\vspace{-1cm}
{\bf Extended Data Figure 5} -- Time evolution of the two-component model of Nova Del 2013.  The model consists of a circular core surrounded by a halo ring.  The expansion rate and size ratio were determined by minimizing $\chi^2$ across all of the nights while allowing the flux ratio to vary night by night.  In plotting the images, we used the flux ratio measured directly for the first three nights, the linear fit plotted in Figure~3 of the main paper for $t$ = 4--27 days, and our measurement that the ring contributes an average of 68\% of the light on the last five nights (dust emission in the outer layers).  Each panel is 12 mas on a side.  We scaled the model flux by the infrared magnitude measured on each night to show how the surface brightness changes.  The time since the explosion (in days) is indicated in each panel. Intensity refers to the flux per unit area.

\end{document}